# Miles Per Gallon: Efficiency Metrics & Preliminary Benchmarking Results for Very Efficient Low Latency Speech Enhancement

**Andrew J.R. Simpson**
magichearing.ai
London, UK
andrew@magichearing.ai

**Fabio Ritrovato**
magichearing.ai
London, UK
fabio@magichearing.ai

## Abstract

In automotive engineering, Miles Per Gallon (MPG) captures the economic efficiency of a car and BHP-Per-Tonne captures dynamic performance, providing interpretable and indispensable ratios. Here, we introduce corresponding efficiency ratio metrics for speech enhancement, leveraging parameter count and latency respectively. These intuitive *efficiency metrics* capture the various trade-offs in speech enhancement, thereby elucidating (and re-arranging) benchmark rankings. In the process, we briefly introduce a Very Efficient Low Latency Speech Enhancement (VELLSE) model designed according to these efficiency metrics. We demonstrate state-of-the-art results on the VoiceBank-DEMAND benchmark across both *lightweight* and *heavyweight* classes of model.

## 1 Introduction

Real-time single-channel speech enhancement has a wide range of applications [1-3] including consumer electronics, hearing aids, Telecoms and as a front-end for ASR systems.

Theoretical constraints are dominated by algorithmic latency for real-time operation [1], practical constraints are dominated by hardware resources (RAM & GOPS/MACS) [3] and functional constraints are dominated by perception of latency and sound quality [2].

Because latency is determined by I/O buffer size and processing time, the performance of most methods is proportional to latency [1, 3]. Similarly, consistent with neural networks in general, the performance of most methods is proportional to model size [1, 2]. This yields a fundamental trade-off in each case.

To illustrate this trade-off, in Fig. 1 we replot results from Fig. 4 of [1], where UTMOS scores are given as a function of latency for the Real-Time RE-USE model (with look-ahead=0) at various exit layers (where each successive layer of computation imposes an additional latency penalty).

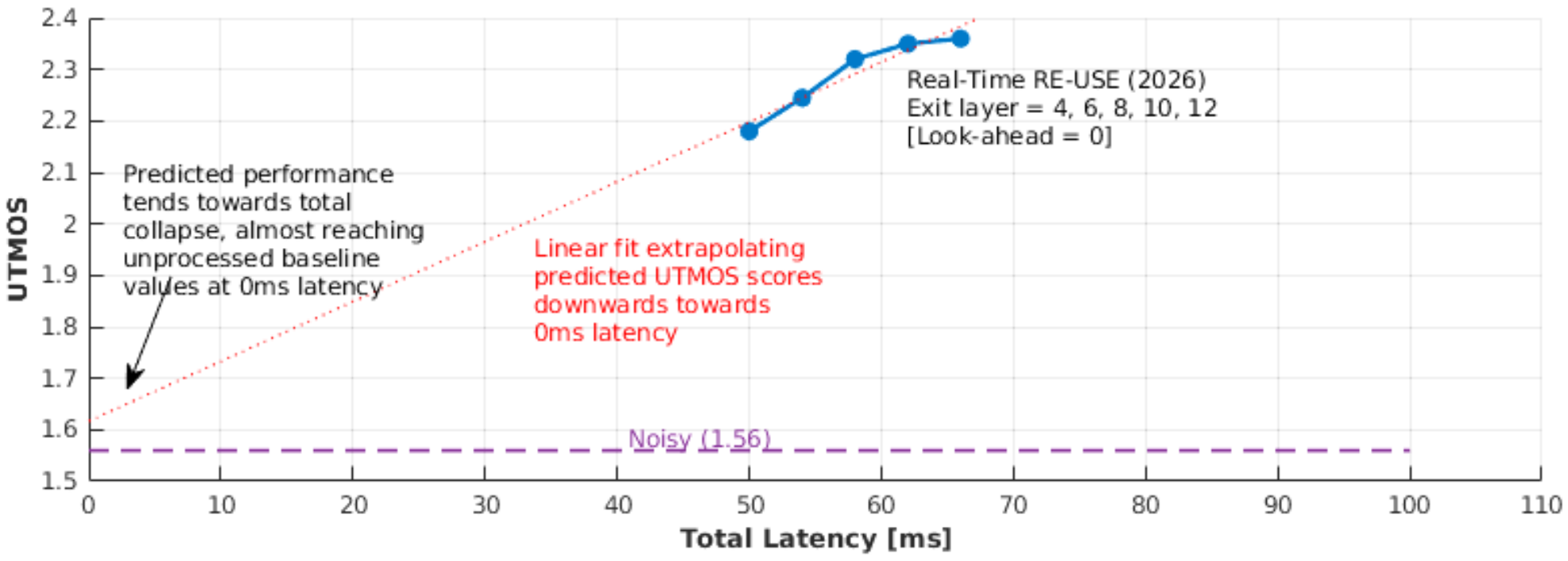

**Figure 1: Illustration of the trade-off between performance and latency.** Data replotted from Fig. 4 of [1] with linear fit to estimate performance losses incurred with decreasing latency.

In Fig. 1 we apply a linear fit, allowing us to crudely estimate the potential performance penalty that might be incurred should the model be capable of being scaled downwards to obtain a latency of, say, 3ms (i.e., comfortably below the threshold of perception). In this particular case, towards 0ms latency output UTMOS scores are predicted to almost intersect the unprocessed level (indicated by the dashed line). A similar trend is evident if the same results are analysed as a function of look-ahead (see Fig. 4 of [1]), where look-ahead increases performance at the expense of increased latency.

### 1.1 Efficiency metrics: motivation and intuitions

For speech enhancement, although several possible efficiency metrics are conceivable, we particularly choose to address the most immediately useful: parameter count and algorithmic latency. Parameter count captures economic efficiency and latency captures dynamic efficiency (i.e., as comparable to MPG and BHP-per-tonne respectively).

Generally, efficiency ratio measures balance the *quality* of the outcome against the *cost* of the outcome. Here, for speech enhancement, outcome quality is measured in terms of output signal error/distortion metrics and hence each such metric may be qualified as a ratio according to the respective costs. In this case, the relevant costs are parameter count and latency. Namely, we end up with ratios of the form *outcome-quality-per-parameter* or *outcome-quality-per-second*.

Parameter count efficiency allows us to understand the fundamental functional intelligence of the system weighed against its cost in machine learning terms. Latency efficiency allows us to incorporate the perceptual significance of latency, penalising late-arriving sound.

Here, for the sake of convenience, in the denominator, we use units of *million parameters* (Mp) for size and *milliseconds* (ms) for latency. In the numerator, we take the metrics associated with the VoiceBank-DEMAND benchmark as in [1,3].

Next, we give an overview of the benchmark test methodology, the metrics, some training caveats, some basic non-parametric statistics and we list the cohort of models included in each benchmarking class. Finally, we briefly introduce our new model and place it in the rankings for each class.

## 2 Methodology

We evaluated performance on the VoiceBank-DEMAND benchmark [4], for convenient comparison to other well-known methods, according to well established 'intrusive' metrics: SI-SDR [5], PESQ [6,7], STOI/ESTOI [8]. However, because of how our model operates in the time-domain, we substitute the original SDR of BSSEval 3.0 for SISDR because our time-domain model is not STFT-based and has effectively arbitrary phase (a consequence of it's very low latency operation). We are aware that SDR tends to over-estimate SISDR values on this particular benchmark (by around 0.5dB according to [5]) but is otherwise comparable. The other metrics (PESQ, ESTOI/STOI), which essentially measure sub-band envelope correlation, are more appropriate for our model.

### 2.1 Brief discussion of metrics & interpretation

With our target of real-time human perception, we optimise our model for PESQ (and for latency) and let the other metrics fall where they may. PESQ has the advantage of being based on auditory modelling, including equal-loudness-contour-shaped perceptual weighting of sensitivity as a function of frequency. PESQ is therefore inherently *speech-focused* simply by virtue of the famous 'speech range' most commonly characterised in the equal loudness contours of Fletcher & Munson [9].

Critically, in contrast to SISDR, SDR, STOI/ESTOI, this metric is insensitive to, for example, subsonic and ultrasonic energy (and/or envelope modulations thereof). It is also insensitive to carrier-frequency phase shift/drift.

Consistent with the speech-range of the equal loudness contours, energy in speech itself is very specifically distributed over frequency - the sensitivity of the ear matches well the energy distribution of speech. Hence, metrics which are agnostic to this sensitivity are liable to misrepresent the error signal as far as it concerns human auditory perception.

With this in mind, we also report STOI/ESTOI/SDR calculated after applying a 100Hz high-pass filter to the clean reference signals (where our model name is appended in the results tables with a '[HPF]'). The high-pass filter at 100Hz does not remove any speech energy from the clean reference signals but it does remove inadvertent subsonic noise from the reference target signals which, though barely audible to a human listener, contributes substantial power by virtue of its low frequency nature. Note that PESQ measurements do not much change in this case because the auditory filter bank already accounts for the inaudibility of this energy, but the perception-agnostic metrics change their readings considerably.

## 2.2 Heavyweight and Lightweight models

Our benchmarking analysis is undertaken in two classes of models: *Heavyweight* and *Lightweight*. The *Heavyweight* class addresses the mostly *generative* class of models in [1]. The class leader here only runs real-time on an Nvidia A100 GPU with ~20TFLOPS (and only at exit layer 8 or below). The *Lightweight* class addresses the more economical models of [3], targeted at hearing aids and wireless earbud SoCs in the range of GFLOPS. Our model is compared in both contexts, according to the specific metrics used by convention in each class.

All reported SDR measures are made at 48kHz using the original 48kHz VoiceBank-DEMAND test set. For compatibility with the conventions of [1] (i.e., *Heavyweight class*), we accompany wide-band ESTOI measurements with 16kHz PESQ measurements. For compatibility with the conventions of [3] (i.e., *Lightweight class*), STOI is used and PESQ is measured at 8kHz. We include STOI/ESTOI/SDR [HPF] metrics for reference as representing 'cleaner clean' reference speech signals.

For each output metric, we calculate two respective efficiency metrics (for parameter count and latency respectively). For latency, we consider algorithmic latency only (by convention).

We do not attempt to cross-compare the two classes of models because of the incompatibility of the metrics used in each case and because of the specialised purpose of each group.

## 2.3 VELLSE model in brief

Our very-efficient-low-latency-speech-enhancement (**VELLSE**) architecture is designed to run on-the-ear on the Qualcomm S5/S7 TWS SoC platform. This imposes constraints of ~3/6mb RAM model size (i.e., 1.5/3m params in FP16), ~64 GOP/s NPU-driven compute limit and the perceptual limit of latency <6ms (including computational contribution). Our single-channel time-domain architecture has 1.5m parameters and operates natively at 48kHz, rendering clean speech at the output with algorithmic latency of 1.98ms. Practically speaking, our 1.98ms algorithmic latency may be accompanied (i.e., during firmware implementation) by a similar 2-3ms computational+buffering latency overhead to combine for a total latency that is below the 6ms detection threshold for human auditory perception.

## 2.2 Training caveats

Our architecture is trained on a proprietary 48kHz training set featuring approximately 2hrs of speech and we optimized for output signal integrity and PESQ. In keeping with recent convention [1], we did not train on VoiceBank-DEMAND data.

## 2.3 Benchmark cohort

We include as *Heavyweight class* models those models addressed in [1]: Diffusion Buffer (2025) [10], DEMUCS (2020) [11], DeepFilterNet3 (2023) [12], Stream.FM (2025) [13], Real-time RE-USE (2026) [1].

We include as *Lightweight class* models those models addressed in [3]: RNNoise (2018) [14], PercepNet (2020) [15], FullSubNet (2021) [16], DeepFilterNet (2022) [17], GaGNet (2022) [18], Fast FullSubNet (2023) [19], GTCRN (2024) [20], LiSenNet (2024) [21], UL-UNAS (2026) [3] and we include a few variants thereof trained with PESQ loss functions. All results (except our own) are given as reported in [1] and [3].

## 2.4 Non-parametric statistical methods: permutation tests of ranking

In the tables of results, **bold-face** indicates the first-ranked model in any given category (i.e., column) and other ranks may be ascertained by inspection. Asterisks (*) denote statistical significance of rank position ($p<0.01$ or better) according to a one-tailed permutation test at 95 percent sub-sample-rate for $N$=10,000, (with Bonferroni correction for multiple tests) inclusive of rounding to decimal places (by convention as in respective tables in [1,3]). In other words, for each permutation test, the test-set metric distribution in question was randomly resampled without duplication, at a 95% sub-sampling rate, 10,000 times, with each sub-sample distribution then yielding a sub-sample mean that is then ranked against the table and tested to count what proportion of times it fails to meet-or-exceed the value necessary for its ranking given in the table. This yields, by proportion of the 10,000 comparisons, a density *p-value*. Essentially, this tests to see how reliable the claimed ranking is with respect to the idiosyncrasies of the test set (and/or metrics): a small (i.e., *significant*) *p-value* indicates that even when the test-set output distributions are sampled with a 5% random-omission rate the same ranking or better still reliably occurs. This seems a prudent mitigation against spurious results given the convention of using mean-averages to capture such results despite them showing obvious near-ceiling effects (and on fairly arbitrary input distributions).

## 3 Results

Table 1 shows results from our model appended to those of the *Heavyweight class* of models [1] for comparison. Table 2 shows similar results for the *Lightweight class* of models [3]. We include latency, sample rate & parameter counts for reference (noting upper sample rate where several options are available for a given model).

**Table 1: Benchmark results (*Heavyweight class)*** for various *Heavyweight class* models on the VoiceBank-DEMAND benchmark. Asterisk (*) indicates statistical significance of ranking (see section 3). $L_{algo}$ refers to algorithmic latency and Params refers to millions of model parameters. PESQ is 16kHz and SDR instead of SISDR for our entries.

| Model | PESQ | ESTOI | SISDR | $L_{algo}$[ms] | Fs[kHz] | Params |
|---|---|---|---|---|---|---|
| Noisy Input | 1.97 | 0.79 | 8.4 | - | - | - |
| Diffusion Buffer (2025) | 2.45 | 0.84 | 14.5 | 176 | 16 | 2.2M |
| DEMUCS (2020) | 2.60 | 0.85 | 15.1 | 41 | 16 | 33.5M |
| DeepFilterNet3 (2023) | 2.71 | 0.84 | 17.3 | 40 | 48 | 2.14M |
| Stream.FM (2025) | 2.72 | 0.85 | 13.4 | 32 | 16 | 52.5M |
| Real-time RE-USE (2026)$_{(lookahead=0)}$ | 2.76 | **0.86** | 18.6 | 40 | 48 | 2.9M |
| Real-time RE-USE (2026)$_{(lookahead=1)}$ | **2.82** | **0.86** | 18.8 | 60 | 48 | 2.9M |
| **VELLSE** v1.0$_{alpha}$ | 2.51* | 0.78 | 16.0* | **1.98** | 48 | **1.5M** |
| **VELLSE** v1.0$_{alpha\ [HPF]}$ | 2.51* | **0.90*** | 16.5* | **1.98** | 48 | **1.5M** |

**Table 2: Benchmark results (*Lightweight class)*** for various *Lightweight class* models. Asterisk (*) indicates statistical significance of ranking (see section 3). $L_{algo}$ refers to algorithmic latency and Params refers to millions of model parameters. PESQ is 8kHz. Note, STOI instead of ESTOI by convention and SDR instead of SISDR for our entries.

| Model | PESQ$_{[8khz]}$ | STOI | SISDR | $L_{algo}$[ms] | Fs[kHz] | Params |
|---|---|---|---|---|---|---|
| Noisy Input | 1.97 | 0.921 | 8.4 | - | - | - |
| RNNoise (2018) | 2.29 | - | - | 30 | 48 | 0.06M |
| PercepNet (2020) | 2.73 | - | - | 30 | 48 | 8M |
| FullSubNet (2021) | 2.89 | 0.94 | - | 80 | 16 | 5.64M |
| DeepFilterNet (2022) | 2.81 | 0.94 | 16.6 | 40 | 48 | 1.8M |
| GaGNet (2022) | 2.79 | 0.93 | - | 30 | 16 | 5.95M |
| Fast FullSubNet (2021) | 2.79 | 0.93 | - | 80 | 16 | 6.84M |
| GTCRN (2024) | 2.87 | 0.94 | - | 40 | 8 | 0.05M |
| LiSenNet (2024) | 2.95 | 0.94 | - | 48 | 16 | **0.04M** |
| LiSenNetb (2024)$_{w/pesq\ loss}$ | 3.07 | 0.94 | - | 48 | 16 | **0.04M** |
| UL-UNAS (2026) | 2.96 | 0.94 | 18.2 | 48 | 16 | 0.17M |
| UL-UNAS (2026)$_{w/pesq\ loss}$ | 3.09 | 0.94 | 18.5 | 48 | 16 | 0.17M |
| **VELLSE** v1.0$_{alpha}$ | **3.18*** | 0.91 | 16.0 | **1.98** | 48 | 1.5M |
| **VELLSE** v1.0$_{alpha\ [HPF]}$ | 3.16* | **0.96*** | 16.5 | **1.98** | 48 | 1.5M |

### 3.1 *Heavyweight class* models

For PESQ, our rank placement between Diffusion Buffer and DEMUCS is statistically significant (*p<0.01*). For ESTOI, our ranking is below that of the noisy input mixtures, but for ESTOI [HPF], our ESTOI value of 0.9 provides a very different ranking (*p<0.01*). Our SDR / SISDR value of 16/16.5dB places us ahead of Diffusion Buffer & DEMUCS & Stream.FM but behind Real-time RE-USE &

DeepFilterNet3.

Our latency of 1.98ms is almost 1/90th that of Diffusion Buffer and our parameter count of 1.5M parameters is less than 1/20th that of DEMUCS.

### 3.2 *Lightweight class* models

In Table 2, the first-place ranking of our PESQ result is statistically significant (*p<0.01*). Our STOI result is poor but our STOI [HPF] result is ranked first (*p<0.01*), illustrating the effect of sub-sonic accidental noise in the reference clean speech signals. Our latency of 1.98ms remains comfortably below that of the field but, in this context, our parameter count is not close to top-ranked (smallest).

### 3.3 Efficiency Analysis: MPG & BHP-Per-Tonne

Tables 3 & 5 provide benchmark results for parameter count efficiency for *Heavyweight class* & *Lightweight class* models respectively, relating PESQ, SISDR & STOI/ESTOI numerators to parameter count denominator respectively.

Tables 4 & 6 provide benchmark results for latency efficiency for *Heavyweight class* & *Lightweight class* models respectively, relating PESQ, SISDR & STOI/ESTOI numerators to ms latency denominator respectively.

#### *3.2.1 Heavyweight class efficiency metrics*

In Table 3, our formerly mid-ranked PESQ, ESTOI & SDR scores (of Table 1) are transformed into top-ranked placement by virtue of size efficiency. Other aspects of the down-table rankings have also shifted considerably, particularly the improved ranking of DeepFilterNet owing to very modest parameter count.

In Table 4, our model is ranked first across all latency efficiency metrics, capturing the large leverage of our very low latency in terms of an order of magnitude difference. In the extreme case, our latency efficiency metric ratios are around two orders of magnitude larger than those of Diffusion Buffer.

#### *3.2.2 Lightweight class efficiency metrics*

In Table 5, by virtue of our relatively (in this context) large model, our model is penalised into mid-table rankings across all metrics. Beyond the immediate concerns of our model falling in the rankings here, some of the other upper rankings have shifted or reversed (versus Table 2).

However, in Table 6 our model is again comfortably ranked first across all latency efficiency metrics, capturing the large leverage of our very low latency in terms of one or even two orders of magnitude advantage.

**Table 3: Size Efficiency *(Heavyweight class)*** for various *Heavyweight* class models for VoiceBank-DEMAND test set as ratio of the number of parameters in each respective model. Mp refers to million model parameters. Asterisk (*) indicates statistical significance at *p<0.01* (*Bonferroni-corrected, one-tailed permutation test at 95% interval*). PESQ is 16kHz, SDR instead of SISDR for our entries.

| **Model** | **PESQ** [MOS/Mp] | **ESTOI** [r/Mp] | **SISDR** [dB/Mp] |
|---|---|---|---|
| Diffusion Buffer (2025) | 0.110 | 0.038 | 0.653 |
| DEMUCS (2020) | 0.078 | 0.025 | 0.451 |
| DeepFilterNet3 (2023) | 1.266 | 0.392 | 8.084 |
| Stream.FM (2025) | 0.052 | 0.016 | 0.257 |
| Real-time RE-USE (2026) (lookahead=0) | 0.952 | 0.296 | 6.414 |
| Real-time RE-USE (2026) (lookahead=1) | 0.972 | 0.296 | 6.483 |
| **VELLSE** v1.0$_{alpha}$ | **1.673*** | **0.52*** | **10.67*** |
| **VELLSE** v1.0$_{alpha\ [HPF]}$ | **1.673*** | **0.60*** | **11.00*** |

**Table 4: Latency Efficiency** ***(Heavyweight class)*** for various *Heavyweight* class models for VoiceBank-DEMAND test set as ratio of the algorithmic (only) latency of each respective model. Ms refers to milliseconds. Asterisk (*) indicates statistical rank significance at *p<0.01* (*Bonferroni-corrected, one-tailed permutation test at 95% interval*). PESQ is 16kHz, SDR instead of SISDR for our entries. HPF denotes our testing using

| **Model** | **PESQ** [MOS/ms] | **ESTOI** [r/ms] | **SISDR** [dB/ms] |
|---|---|---|---|
| Diffusion Buffer (2025) | 0.014 | 0.005 | 0.082 |
| DEMUCS (2020) | 0.063 | 0.021 | 0.368 |
| DeepFilterNet3 (2023) | 0.068 | 0.021 | 0.432 |
| Stream.FM (2025) | 0.085 | 0.026 | 0.419 |
| Real-time RE-USE (2026) (lookahead=0) | 0.069 | 0.021 | 0.465 |
| Real-time RE-USE (2026) (lookahead=1) | 0.047 | 0.014 | 0.313 |
| **VELLSE** v1.0alpha | **1.268*** | **0.394*** | **8.081*** |
| **VELLSE** v1.0alpha [HPF] | **1.268*** | **0.454*** | **8.333*** |

**Table 5: Size Efficiency** ***(Lightweight class)*** for various *Lightweight* class models for VoiceBank-DEMAND test set as ratio of the number of parameters in each respective model. Mp refers to million model parameters. PESQ is 8Hz. Note: STOI not ESTOI.

| **Model** | **PESQ[8khz][MOS/Mp]** | **STOI [r/Mp]** | **SISDR [dB/Mp]** |
|---|---|---|---|
| RNNoise (2018) | 38.2 | - | - |
| PercepNet (2020) | 0.34 | - | - |
| FullSubNet (2021) | 0.5 | 0.17 | - |
| DeepFilterNet (2022) | 1.61 | 0.52 | 9.24 |
| GaGNet (2022) | 0.48 | 0.16 | - |
| Fast FullSubNet (2021) | 0.43 | 0.14 | - |
| GTCRN (2024) | 57.4 | 18.8 | - |
| LiSenNet (2024) | 73.8 | 23.4 | - |
| LiSenNetb (2024) w/pesq loss | **76.8** | **23.5** | - |
| UL-UNAS (2026) | 17.4 | 5.5 | 107.3 |
| UL-UNAS (2026) w/pesq loss | 18.2 | 5.5 | **108.7** |
| **VELLSE** v1.0alpha | 2.12 | 0.61 | 10.67* |
| **VELLSE** v1.0alpha [HPF] | 2.11 | 0.64 | 11.00* |

**Table 6: Latency Efficiency *(Lightweight class)*** for various *Lightweight* class models for VoiceBank-DEMAND test set as ratio of the algorithmic (only) latency of each respective model. Ms refers to milliseconds. Asterisk (*) indicates statistical rank significance at *p<0.01* (*Bonferroni-corrected, one-tailed permutation test at 95% interval*). PESQ is 8kHz. Note: STOI not ESTOI.

| Model | $\text{PESQ}_{[8khz]}$[MOS/ms] | STOI [r/ms] | SISDR [dB/ms] |
|---|---|---|---|
| RNNoise (2018) | 0.076 | - | - |
| PercepNet (2020) | 0.091 | - | - |
| FullSubNet (2021) | 0.036 | 0.012 | - |
| DeepFilterNet (2022) | 0.072 | 0.023 | 0.41 |
| GaGNet (2022) | 0.094 | 0.031 | - |
| Fast FullSubNet (2021) | 0.035 | 0.012 | - |
| GTCRN (2024) | 0.071 | 0.023 | - |
| LiSenNet (2024) | 0.061 | 0.019 | - |
| LiSenNetb (2024) w/pesq loss | 0.064 | 0.019 | - |
| UL-UNAS (2026) | 0.062 | 0.020 | 0.38 |
| UL-UNAS (2026) w/pesq loss | 0.064 | 0.020 | 0.38 |
| **VELLSE** v1.0alpha | **1.606*** | **0.460*** | **8.08*** |
| **VELLSE** v1.0alpha [HPF] | **1.596*** | **0.485*** | **8.33*** |

## 4 Conclusions

We have introduced some intuitive, interpretable efficiency metrics for speech enhancement. We have applied them to a well known real-time speech enhancement benchmark for two classes of single-channel speech enhancement methods. We have briefly introduced preliminary results for a new model, that we intend to describe fully at a later date, and have demonstrated state-of-the-art results across both lightweight and heavyweight classes of speech enhancement models.

In latency efficiency, our model is top-ranked across both heavyweight and lightweight classes of model in all metrics.

With more time and resources, we would like to extend our efficiency metrics to include bandwidth efficiency (e.g., performance-per-octave) and training data efficiency. We would also like to extend our efficiency metrics to include computational latency and power consumption.

Training on a single local machine with two TitanX GPUs, and with only ~2hrs of good quality training data, our VELLSE model has shown quite competitive results. In terms of efficiency, our method is state-of-the-art across most measures in both classes. This is the result both of our inherently small latency and our inherently small model size (both driven by our requirement of running on the Qualcomm S5/S7 NPU SoC and without perceptible latency).

With our very limited lab resources (training data & compute), combined with the relatively small size of the current model, there is an obvious anticipation of untapped potential. Our initial experiments in scaling our model upwards indicate typical scaling laws apply (benchmark/metric ceiling effects notwithstanding). Thus, there is obviously much potential to scale training data, model size and compute resources to levels commensurate with the field.

We have yet to test our model as a front-end for ASR but we suspect that the latency advantage might be useful and we have no reason to doubt that the fundamental speech enhancement performance will translate as well as other similarly ranked speech enhancement methods.

Our architecture, as captured here, is running real-time on our laptop-CPU-based demonstrator device (with one independent model for each ear) and can be experienced live in-person (by appointment) combined with ANC-equipped earphones for an ‘enhanced transparency mode’ experience. We expect to have a running port on a Qualcomm S7 TWS reference design as soon as we’re able to obtain the necessary hardware.

## Acknowledgments

We wish to acknowledge the painful electricity costs and other more limiting resource constraints of training for this project. If you - the reader - know of a lab with serious resources that would like to work with us to scale-up our project, please reach out and make us an offer. And if you have a spare S7 reference design just lying around the office in an unopened box, please send it our way.